\documentclass[12pt]{article}
\usepackage{graphicx}                   
\usepackage{amsmath}
\newcommand{\word}[1]{\,\,\mbox{#1}\,\,}
\newcommand{\reff}[1]{(\ref{#1})}                      
\newcommand{\beq}{\begin{equation}}
\newcommand{\eeq}[1]{\label{#1}\end{equation}} 
\newcommand{\beg}{\begin{equation*}}
\newcommand{\eeg}{\end{equation*}}

\newcommand{\expo}[1]{\mbox{e}^{#1}}

\newcommand{\der}[1]{\left(\dfrac{d}{d{#1}}\right)}

\newcommand{\rdn}{r_d(n)}
\newcommand{\sumprime}{\sideset{}{'}\sum}
\newcommand{\bsplit}{\begin{split}}
\newcommand{\esplit}{\end{split}}
\newcommand{\lrack}{\Bigl(}
\newcommand{\rrack}{\Bigr)}
\newcommand{\bino}[2]{\Bigl(\genfrac{}{}{0pt}{}{#1}{#2}\Bigr)}
\begin{document}
\title{Generating Function of the Arithmetical Function $\rdn$ and its Relation to the Casimir Energy}
\author{\thanks{Email: aedery@ubishops.ca} Ariel Edery\\Bishop's University, Physics Department\\Lennoxville, Quebec\\J1M1Z7} 
\date{}

\maketitle

\begin{abstract}
We obtain analytical expressions for the generating function $\xi_d(\lambda)$ of the sum of $d$-squares arithmetical function $\rdn$ where $\lambda$ is a free parameter. The original $d$-dimensional infinite sum is reduced to a formula containing a single finite sum over a convergent series. We compare the formulas to numerical computations and show that the percentage difference is negligible at small $\lambda$ for various values of $d$. $\xi_d(\lambda)$ divides naturally into two terms and we show that one term has a direct physical application to the $d$-dimensional Casimir energy of massless scalar fields in cubic cavities. 
\end{abstract} 

\newpage

\subsection*{\center{I. Introduction}}

The representation of integers as a sum of $d$ squares is one of the classical problems in Number Theory \cite{Hardy}. The number of representations of a positive integer $n$ by $d$ squares, allowing zeros and distinguishing signs and order, is denoted by the arithmetical function $\rdn$ where it is common to write $r(n)$ for $r_2(n)$. Great ingenuity and effort has gone into obtaining explicit formulas for these functions. In particular, a systematic treatment for even values of $d\le 8$ existed by 1940. However, a systematic treatment for odd values with $d\le7$ was only obtained in 2002 by Shimura \cite{Shimura}. Explicit formulas for any value of $d$ do not presently exist. Fortunately, explicit knowledge of the functions is not required for the calculation of generating functions. In this paper we are interested in a particular generating function of the arithmetical function $rd(n)$ because of its connection with a vacuum boundary effect in quantum field theory called the Casimir energy. The most-well known cases of infinite series of $r_d(n)$ are the Dirichlet and power series involving $r(n)$. The Dirichlet series is given by \cite{Hardy},
\beq
\sum_{n=1}^{\infty} \dfrac{r(n)}{n^s} = 4 \,\zeta(s)\,L(s) \word{for} s>1 \word{where} L(s) = 1^{-s}-3^{-s}+5^{-s} - \ldots
\eeq{Dirich}
while the power series is given by \cite{Hardy},
\beq
\sum_{n=1}^{\infty} r(n) \, x^n = 4 \lrack \dfrac{x}{1-x} - \dfrac{x^3}{1-x^3} + \dfrac{x^5}{1-x^5} - \ldots\rrack.
\eeq{pasteq}
An important result is that the average order of $r(n)$ is simply $\pi$:  
\beq
\lim_{N \to \infty} \dfrac{1}{N} \sum_{n=0}^{N} r(n) = \pi.
\eeq{yup}
There is a clear geometrical interpretation of the above result. The sum up to $N$ is equivalent to counting the lattice points inside and on a circle given by $x^2+y^2 = r^2$ where the radius $r= \sqrt{N}$. If each lattice point is assigned an area of one unit, the total area approaches the area $\pi r^2$ and hence $\pi\,N$ as $r$ approaches infinity. Good estimates of the correction to $\pi\,N$ for finite $N$ have been obtained as well as the distribution of the lattice points on the circle \cite{Cilleruello}. 

In this paper, we are interested in results where the dimension $d$ is not fixed a-priori. Specifically, we obtain accurate formulas for the infinite series of the arithmetical function $r_d(n)$ multiplied by an exponential damping term raised to the square root of $n$. The generating function is:
\beq
\xi_d(\lambda) \equiv \sum_{n=0}^{\infty} \rdn\,\expo{-\lambda\,\sqrt{n}} 
\eeq{rdn} 
where the free parameter $\lambda$ is a positive real number and yields a convergent series. $\xi_d(\lambda)$ is a function of both the parameter $\lambda$ and the dimension $d$. Note that it deviates from usual power series like \reff{pasteq} because the exponential is raised to the square root of $n$ instead of $n$. The $\sqrt{n}$ is chosen for physical reasons. The frequency of normal modes of a massless scalar field in a $d$-dimensional cubic cavity is proportional to $\sqrt{n}$ (where $n = n_1^2 +n_2^2 + \cdot + n_d^2$ and $n_i$ is the mode in the $i$th direction). These can then be summed in a particular fashion to obtain the $d$-dimensional Casimir energy. Sums over the $\sqrt{n}$ can be obtained from \reff{rdn} by taking the derivative with respect to $-\lambda$. More generally 
\beq
\sum_{n=0}^{\infty} \rdn \,n^{p/2}\,\expo{-\lambda\,\sqrt{n}} = (-1)^p \,\der{\lambda}^{p} \xi_d(\lambda)\,
\eeq{newsum}
where $p$ is a non-negative integer. Hence, with $p=1$, we have a sum over the $\sqrt{n}$ and a possible connection with the Casimir energy in cubic cavities. 

The formula we obtain for $\xi_d(\lambda)$ divides naturally into two terms: an integral term $I_d(\lambda)$ and a second term which we label $C_d(\lambda)$. $I_d(\lambda)$ is equivalent to replacing the multiple sums in $\xi_d(\lambda)$ by multiple integrals. For small $\lambda$, $I_d(\lambda)$ can be viewed as a first approximation to $\xi_d(\lambda)$ with $C_d(\lambda)$ a small correction. However, at large $\lambda$, $C_d(\lambda)$ actually dominates over $I_d(\lambda)$. We compare our formulas to direct numerical computations for values of $d$ ranging from $2$ to $5$ and for $\lambda$ ranging from $0.1$ to $10$. The difference (error) is due to a remainder term we neglect in the derivation of our formulas. For small $\lambda$, the agreement is excellent. For $\lambda \le 1$ we obtain less than $0.02\%$ error and for $\lambda \le 5$ we obtain less than $6.5\%$ error. The percentage difference increases as $\lambda$ and $d$ increases. At $\lambda =10$ the error is $10\%$ at $d=2$ and $32\%$ at $d=5$.  

The remarkable thing is that the formula for $\xi_d(\lambda)$ is already in a form that contains the necessary pieces for a clear formulation of the Casimir energy. In the end, all the information we need is contained in $C_d(\lambda)$. We show how one can obtain the $d$-dimensional Casimir energy of massless scalar fields in cubic cavities by taking the appropriate derivatives and limits of the parameter $\lambda$ in $C_d(\lambda)$.

\subsection*{\center{II. Formulas Obtained}}

In this section we are interested in obtaining analytical expressions for the generating function
\beq
\xi_d(\lambda)\equiv \sum_{n=1}^{\infty} \rdn\,\expo{-\lambda\,\sqrt{n}}\,.
\eeq{rdndef}
The above definition is equivalent to 
\beq 
\xi_d(\lambda)= -1 + \sum_{n_d=-\infty}^{\infty}\cdots\sum_{n_1=-\infty}^{\infty}
\expo{-\lambda\,\sqrt{n_1^2 +n_2^2 + \cdots + n_d^2}} \,.
\eeq{start}
Define the following short-hand form for a $j$-dimensional multiple sum:
\beq
\Bigl(\,\sum\,\Bigr)^j \equiv \sum_{n_j=-\infty}^{\infty}\cdots\sum_{n_1=-\infty}^{\infty}
\expo{-\lambda\,\sqrt{n_1^2 +n_2^2 + \cdots + n_j^2}}\,.
\eeq{multiple}
A sum from minus to plus infinity that excludes zero is labeled $\sumprime$. It is convenient to construct the following $(j+1)$-dimensional sum:
\beq
\Lambda_j(\lambda) \equiv \sumprime\Bigl(\,\sum\,\Bigr)^j =\sumprime_{n=-\infty}^{\infty}\,\,\sum_{n_j=-\infty}^{\infty}\cdots\sum_{n_1=-\infty}^{\infty}
\expo{-\lambda\,\sqrt{n^2+ n_1^2 +n_2^2 + \cdots + n_j^2}}\,.
\eeq{j1}
Note that the last sum in $\Lambda_j(\lambda)$, the sum over the variable labeled $n$, does not contain a zero. The $d$-dimensional sum in \reff{start} can now be expanded in the following fashion:
\beq
\begin{split}
&\sum_{n_d=-\infty}^{\infty}\cdots\sum_{n_1=-\infty}^{\infty}
\expo{-\lambda\,\sqrt{n_1^2 +n_2^2 + \cdots + n_d^2}} \\
&= 1 + \sumprime + \sumprime\,\sum + \sumprime\,\Bigl(\,\sum\,\Bigr)^2 +\cdots + \sumprime\,\Bigl(\,\sum\,\Bigr)^{d-1} \\
&= 1 + \sum_{j=0}^{d-1} \sumprime\,\bigl(\,\sum\,\bigr)^j \\
&= 1 + \sum_{j=0}^{d-1} \Lambda_j(\lambda)
\end{split}
\eeq{testr}
Substituting \reff{testr} into \reff{start} yields  
\beq 
\xi_d(\lambda) = \sum_{j=0}^{d-1} \Lambda_j(\lambda)\,.
\eeq{def}
The goal now is to find analytical expressions for $\Lambda_j(\lambda)$. To this end we apply the Euler-Maclaurin formula that converts sums to integrals. The Euler-Maclaurin formula is \cite{Arfken}:  
\beq
\sum_{n=1}^{\infty} f(n) =\int_0^\infty f(x)\,dx - \dfrac{1}{2}\,f(0)
+ \sum_{p=1}^{q}\dfrac{1}{(2p)!}\,B_{2p}\,f^{(2p-1)}(0) + R_q
\eeq{euler}
where $f^{(2p-1)}(0)$ are odd derivatives evaluated at zero and $q$ is a positive integer. $R_q$ is the remainder term given by \cite{Arfken}
\beq
R_q= -\dfrac{1}{(2q)!} \int_0^1\, B_{2q}(x) \,\sum_{\nu=0}^{\infty}\, f^{2q}\,(x+\nu)\,dx \,. 
\eeq{remainder}
where $B_{2q}(x)$ are Bernoulli functions. We will neglect the remainder term $R_q$ in deriving formulas for $\xi_d(\lambda)$. The error introduced by omitting $R_q$ is calculated in the section where we compare the formulas to numerical computations. 

We want to make use of the Euler-Maclaurin formula to determine $\Lambda_j(\lambda)$. The function $f$ being summed is the exponential function in \reff{j1}. For this function, $f^{2p-1}(0)$ is zero for the first $j$ sums in \reff{j1} (all sums except the last one). A proof of this is given in \cite{Ariel}. When $f^{2p-1}(0)=0$, the sum from $p=1$ to $q$ in\reff{euler} is zero independent of $q$. This implies that $R_q$ given by \reff{remainder} must be independent of $q$ for the case of our exponential function. We prove this explicitly in Appendix A. Note that if $R_q$ is nonzero, it has the same nonzero value for all $q$ even though $1/(2q)!$ approaches zero as $q \to \infty$. 

With $f^{2p-1}(0)=0$ for the first $j$ sums we neglect $R_q$ and  the Euler-Maclaurin formula \reff{euler} for the first $j$ sums reduces to
\beq
\sum_{n=1}^{\infty} f(n) =\int_0^\infty f(x)\,dx - \dfrac{1}{2}\,f(0) \,.
\eeq{Euler2}
The function $f$ in \reff{j1} has the property $f(n_i)=f(-n_i)$. The sum over a given $n_i$ can therefore be written as 
\beq
\begin{split}
\sum_{n_i=-\infty}^{\infty} f(n_i) &= 2\,\sum_{n_i=1}^{\infty}\,f(n_i) + f(0)\\
&=2 \Biggl(\int_0^\infty f(x)\,dx - \dfrac{1}{2}\,f(0)\Biggr) + f(0)\\
&=\int_{-\infty}^{\infty} f(x)\,dx 
\end{split}
\eeq{firstsum} 
where we used \reff{Euler2}. From \reff{firstsum} we see that each sum in \reff{j1}, except the last one, can be replaced by an integral. This yields 
\beq
\begin{split}
\Lambda_j(\lambda)&= \sumprime\Bigl(\,\sum\,\Bigr)^j \to \sumprime \int^j\\
& = \sumprime_{n=-\infty}^{\infty}\,\int_{-\infty}^{\infty}
\expo{-\lambda\,\sqrt{n^2+ x_1^2 +x_2^2 + \cdots + x_j^2}} dx_1\,dx_2\ldots dx_j \\
&= 2^{j+1}\,\sum_{n=1}^{\infty}\, \int_{0}^{\infty} \expo{-\lambda\,\sqrt{n^2+ x_1^2 +x_2^2 + \cdots + x_j^2}} dx_1\,dx_2\ldots dx_j 
\end{split} 
\eeq{dxj} 
The integral can be expressed in terms of a modified Bessel function \cite{Gradshteyn}:
\beq
\begin{split}
&\int_0^{\infty} \expo{-\lambda\,\sqrt{n^2 +x_1^2 + \cdots + x_j^2}} dx_1\ldots dx_j \\
& = -2^{\frac{1-j}{2}}\,\pi^{\frac{j-1}{2}}\,\dfrac{d}{d\lambda}\Bigl(K_{\frac{j-1}{2}}(\lambda\,n)\,
\Bigl(\frac{n}{\lambda}\Bigr)^{\frac{j-1}{2}}\Bigr)\,.
\end{split}
\eeq{in}
The modified Bessel function $K_{\frac{j-1}{2}}(\lambda\,n)$ has the following identity \cite{Gradshteyn}:
\beq
K_{\frac{j-1}{2}}(\lambda\,n)\,\Bigl(\frac{n}{\lambda}\Bigr)^{\frac{j-1}{2}} = (-1)^{\frac{1-j}{2}}
\Bigl(\frac{d}{\lambda\,d\lambda}\Bigr)^{\frac{j-1}{2}} K_0(\lambda\,n)\,.
\eeq{bess}
Substituting \reff{bess} and \reff{in} into \reff{dxj} yields
\beq
\Lambda_j(\lambda) =  2^{\frac{j+3}{2}}\,\pi^{\frac{j-1}{2}} (-1)^{\frac{3-j}{2}}\dfrac{d}{d\lambda}
\Bigl(\frac{d}{\lambda\,d\lambda}\Bigr)^{\frac{j-1}{2}} \sum_{n=1}^\infty K_0(\lambda\,n)\,.
\eeq{sumko}
The infinite sum over the modified Bessel function $K_0(\lambda\,n)$ has the following series expansion \cite{Gradshteyn}:
\beq
\sum_{n=1}^\infty K_0(\lambda\,n) = \dfrac{1}{2}\left\{C + \ln(\lambda/4\pi)\right\} + \dfrac{\pi}{2\,\lambda} +
\pi \sum_{m=1}^\infty \left(\dfrac{1}{\sqrt{\lambda^2 + 4\,m^2\,\pi^2}} - \dfrac{1}{2\,m\,\pi}\right)\,.
\eeq{series}
Substituting \reff{series} into \reff{sumko} yields
\beq
\begin{split}
\Lambda_j(\lambda)
&= -\dfrac{1}{\lambda^j}\,2^j \,\pi^{\frac{j-1}{2}}\, \Gamma(\tfrac{j+1}{2}) +
\dfrac{1}{\lambda^{j+1}}\,2^{j+1}\,\pi^{\frac{j}{2}}\,\Gamma(\tfrac{j+2}{2}) \\
&\qquad\qquad\qquad\qquad\qquad\qquad+ \lambda \,\,2^{j+2}\,\,\Gamma(\tfrac{j+2}{2})\,\pi^{\frac{j}{2}}\,\chi_j(\lambda)
\end{split}
\eeq{finito}
where
\beq
\chi_j(\lambda) \equiv \sum_{m=1}^\infty \dfrac{1}{(\lambda^2 + 4\,m^2\,\pi^2)^{\frac{j+2}{2}}} \,\,.
\eeq{chi} 
Substituting \reff{finito} into \reff{def} yields:
\beq
\xi_d(\lambda) = \sum_{j=0}^{d-1}\, -\dfrac{2^j}{\lambda^j}\,\,\pi^{\frac{j-1}{2}}\, \Gamma(\tfrac{j+1}{2}) +
\dfrac{2^{j+1}}{\lambda^{j+1}}\,\,\pi^{\frac{j}{2}}\,\Gamma(\tfrac{j+2}{2}) + \lambda \,\,2^{j+2}\,\,\Gamma(\tfrac{j+2}{2})\,\pi^{\frac{j}{2}}\,\chi_j(\lambda)\,.
\eeq{short2}
Three terms are being summed. Note that if the first term is written as $-a_j$, then the second term is $a_{j+1}$. The sum over $j$ of the first two terms is then a telescopic sum equal to $-a_0 + a_d = -1 + \tfrac{2^d}{\lambda^d}\,\,\pi^{\frac{d-1}{2}}\,\, \Gamma(\tfrac{d+1}{2})$. Substituting this result into \reff{short2} yields
\beq
\xi_d(\lambda) = \dfrac{2^d}{\lambda^d}\,\,\pi^{\frac{d-1}{2}}\,\, \Gamma(\tfrac{d+1}{2}) -1 + \lambda \sum_{j=0}^{d-1} 2^{j+2}\,\,\Gamma(\tfrac{j+2}{2})\,\pi^{\frac{j}{2}}\,\chi_j(\lambda)\,.
\eeq{xid} 
Equation \reff{xid} is our final formula for $\xi_d(\lambda)$. The original $d$-dimensional infinite sum \reff{start} has been replaced by a finite sum over the convergent series $\chi_j(\lambda)$ given by \reff{chi}. It is convenient to express \reff{xid} in the form
\beq
\xi_d(\lambda) = I_d(\lambda) -1 + C_d(\lambda)
\eeq{xinew}
where
\beq
I_d(\lambda) \equiv \dfrac{2^d}{\lambda^d}\,\,\pi^{\frac{d-1}{2}}\,\, \Gamma(\tfrac{d+1}{2}) \quad\word{and} \quad C_d(\lambda)\equiv
\lambda \sum_{j=0}^{d-1} 2^{j+2}\,\,\Gamma(\tfrac{j+2}{2})\,\pi^{\frac{j}{2}}\,\chi_j(\lambda) \,.
\eeq{IdCd}
There is a clear interpretation to \reff{xinew}. $I_d(\lambda)$ is identified as the `integral contribution' to $\xi_d(\lambda)$. It is the multiple integral of $\expo{-\lambda\,r}$ over the $d$-dimensional volume where $r=(x_1^2+x_2^2+\cdots+x_d^2)^{1/2}$. This is shown in Appendix B. The expression $I_d(\lambda) -1$ appearing in \reff{xinew} is equivalent to \reff{start} with the $d$ sums replaced by $d$ integrals. This can be viewed as a first approximation to $\xi_d(\lambda)$. $C_d(\lambda)$ is the interesting part for mathematical and physical reasons. Mathematically, $C_d(\lambda)$ is the non-trivial `correction' to the approximation of replacing sums by integrals in \reff{start}. Physically, $C_d(\lambda)$ is the relevant part for the calculation of the Casimir energy as we will demonstrate later. $C_d(\lambda)$ is not necessarily smaller than $I_d(\lambda)$. Their limits as $\lambda \to 0$ and $\lambda \to \infty$ are:
\beq
\begin{split}
C_d(\lambda) & \to 0 \word{and} I_d(\lambda) \to \infty \word{as} \lambda \to 0\\
&\word{and}\\
C_d(\lambda) & \to 1 \word{and} I_d(\lambda) \to 0 \word{as} \lambda \to \infty\,.
\end{split}
\eeq{Cd}
It is clear that $I_d(\lambda)$ dominates at sufficiently small $\lambda$ and $C_d(\lambda)$ dominates at sufficiently large $\lambda$. How small is sufficiently small? For a given dimension, this depends on $\lambda$. Values are quoted in Table 1 for $\lambda$ ranging from $0.1$ to $10$ and $d$ ranging from $2$ to $5$. In this sample, $I_d(\lambda$ dominates for $\lambda=1$ and below and $C_d(\lambda)$ dominates for $\lambda=5$ and greater. It is therefore only valid to view $C_d(\lambda)$ as a correction term for $\lambda$ below a certain value.     
\begin{table}[ht]
\begin{center}
\caption{Values of $C_d(\lambda)$, $I_d(\lambda)$ and their Ratio} 
\includegraphics[scale=0.65]{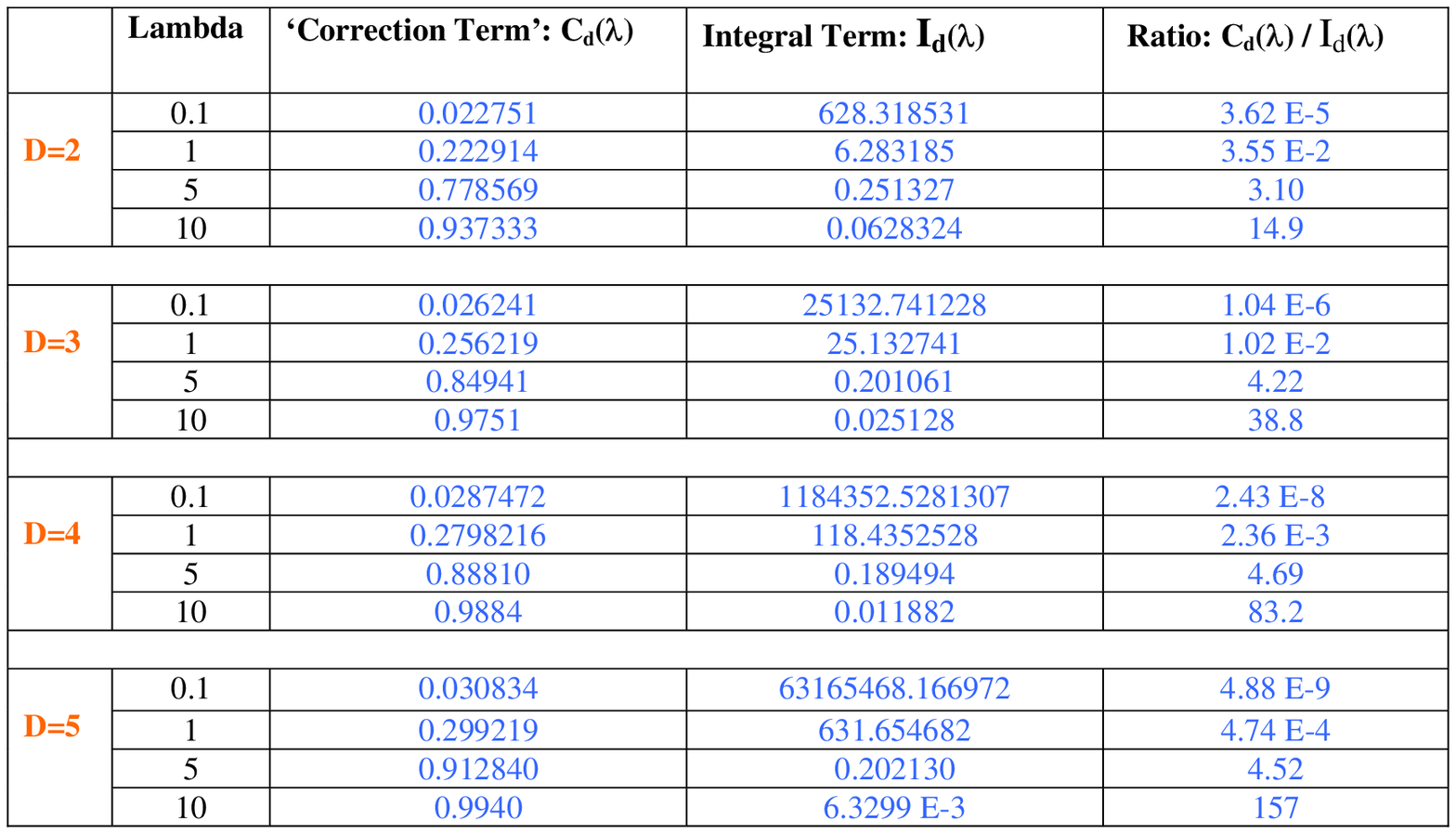}
\end{center}
\end{table}

Formula \reff{xid} together with \reff{chi} for $\chi_j(\lambda)$ are our final formulas needed to determine $\xi_d(\lambda)$. We can go one step further if $0 \!<\!\lambda\!<\!2\pi$. Then $\chi_j(\lambda)$ can be expanded as a power series in $\lambda$ involving Riemann zeta functions. The result is a power series for $\xi_d(\lambda)$ which converges rapidly when $\lambda << 2\,\pi$.  
 
\subsubsection*{Power Series Expansion for $\chi_j(\lambda)$}

For the case $0 \!<\!\lambda\!<\!2\pi$, $\chi_j(\lambda)$ can be expressed as a power series in $\lambda$ with Riemann zeta functions as coefficients. We begin by rewriting \reff{chi} in a form suitable for a binomial expansion:
\beq
\begin{split}
\chi_j(\lambda) & \equiv \sum_{m=1}^\infty \dfrac{1}{(\lambda^2 + 4\,m^2\,\pi^2)^{\frac{j+2}{2}}}\\
&=\dfrac{1}{(2\pi)^{j+2}}\,\sum_{m=1}^\infty \dfrac{1}{m^{j+2}} \Bigl(1 + \dfrac{\lambda^2}{4\,m^2\,\pi^2}\Bigr)^{\frac{-(j+2)}{2}}\,.\\
\end{split}
\eeq{bino}
We can obtain a binomial expansion for $\Bigl(1 + \tfrac{\lambda^2}{4\,m^2\,\pi^2}\Bigr)^{\frac{-(j+2)}{2}}$ if $\lambda^2/4\,m^2\,\pi^2<1$ for all $m$. This is true if $0 \!<\!\lambda\!<\!2\pi$ (the condition $\lambda > 0$ is there because $\lambda$ was assumed from the start to be a positive real number). The binomial expansion yields
\beq
\begin{split}
\Bigl(1 + \dfrac{\lambda^2}{4\,m^2\,\pi^2}\Bigr)^{\frac{-(j+2)}{2}} &= \sum_{n=0}^\infty \bino{\frac{-(j+2)}{2}}{n} \Bigl(\dfrac{\lambda^2}{4\,m^2\,\pi^2}\Bigr)^n\\
&= \sum_{n=0}^\infty \dfrac{(-1)^n}{n!}\,\dfrac{\lambda^{2n}}{4^n\,\pi^{2n}}\dfrac{1}{m^{2n}} \prod_{i=1}^n (i + j/2) \,.
\end{split}
\eeq{nom}
Substituting \reff{nom} into \reff{bino} yields
\beq
\begin{split}
\chi_j(\lambda)&= \dfrac{1}{(2\pi)^{j+2}}\,\sum_{n=0}^\infty \dfrac{(-1)^n}{n!}\,\dfrac{\lambda^{2n}}{4^n\,\pi^{2n}}\prod_{i=1}^n (i + j/2) \sum_{m=1}^\infty \dfrac{1}{m^{j+2 +2n}} \\
&= \dfrac{1}{(2\pi)^{j+2}}\,\sum_{n=0}^\infty \dfrac{(-1)^n}{n!}\,\dfrac{\lambda^{2n}}{4^n\,\pi^{2n}} \zeta(j+2 +2n)\prod_{i=1}^n (i + j/2)\,.\\
\end{split}
\eeq{chichi} 
Expanding \reff{chichi} term by term yields
\beq
\begin{split}
\chi_j(\lambda) &=\dfrac{1}{(2\pi)^{j+2}} \biggl\{\zeta(j+2) - \dfrac{\lambda^2}{4\,\pi^2}\,\zeta(j+4)\,(1+j/2)\\ &\qquad\qquad\qquad\qquad+\dfrac{\lambda^4}{16\,\pi^4\,2!}\,\zeta(j+6)\,(2+j/2)\,(1+j/2) + \ldots \biggr\}\,.
\end{split}
\eeq{finalchi}
Sustituting \reff{finalchi} into \reff{xid} yields
\beq
\begin{split}
\xi_d(\lambda) &= \dfrac{2^d}{\lambda^d}\,\,\pi^{\frac{d-1}{2}}\,\, \Gamma(\tfrac{d+1}{2}) -1 \\
&\qquad\qquad+ \lambda \sum_{j=0}^{d-1} \dfrac{\Gamma(\tfrac{j+2}{2})}{\pi^{j/2 +2}} \biggl\{\zeta(j+2) - \dfrac{\lambda^2}{4\,\pi^2}\,\zeta(j+4)\,(1+j/2)\\
&\qquad\qquad\qquad\qquad+\dfrac{\lambda^4}{16\,\pi^4\,2!}\,\zeta(j+6)\,(2+j/2)\,(1+j/2) + \ldots \biggr\}\,.
\end{split}
\eeq{lambdasmall}
The above power series for $\xi_d(\lambda)$ is valid when $0 \!<\!\lambda\!<\!2\pi$. One of its advantages is that it converges rapidly when $\lambda/2\pi <<1$.

\subsection*{\center{II. Numerical Algorithm and Comparison to Formulas}}

In the process of deriving a formula for $\xi_d(\lambda)$ we omitted the remainder term $R_q$. The formulas will therefore contain some errors. Recall that the exact expression for $\xi_d(\lambda)$ is
\beq
\begin{split}
\xi_d(\lambda) &= \sum_{n=1}^{\infty} \rdn\,\expo{-\lambda\,\sqrt{n}}\, \\
&= -1 + \sum_{n_d=-\infty}^{\infty}\cdots\sum_{n_1=-\infty}^{\infty}
\expo{-\lambda\,\sqrt{n_1^2 +n_2^2 + \cdots + n_d^2}} \,.
\end{split}
\eeq{rdn2} 
We developed an algorithm that computes \reff{rdn2} in a time efficient way by making use of standard permutation formulas to avoid using negative integers and all possible orderings. The algorithm is also constructed with the flexibility of allowing the dimension $d$ to be a variable and user-input. In contrast, if the numbers were generated using $d$-nested for-loops, then
one would need to change by hand the number of for-loops for different dimensions. 

We then calculate $\xi_d(\lambda)$ using the formulas we derived: \reff{xid} together with \reff{chi} 
(it is convenient to use the power series \reff{lambdasmall} for cases when $\lambda$ is small i.e. $\lambda\!<\!<\!2\pi$). We then determine the absolute and percentage difference between the formula and algorithm.
We performed calculations for the following cases: $\lambda = 0.1, 1, 5$ and $10$ and dimensions $d = 2, 3, 4$ and $5$. The results are quoted in table 2. The numbers are quoted to the precision reached with the last digit rounded.

For small $\lambda$, the agreement is excellent. For $\lambda \le 1$ we obtain less than $0.02\%$ error and for $\lambda \le 5$ we obtain less than $6.5\%$ error. It is important to note that the agreement at values like $\lambda=5$ is highly dependent on the contribution made by $C_d(\lambda)$. As previously mentioned, $C_d(\lambda)$ makes a dominant contribution already at $\lambda=5$ (see table 1). The percentage difference increases as $\lambda$ and $d$ increase. For the highest value of $\lambda$ in table 2, $\lambda =10$, the error is $10\%$ at $d=2$ and $32\%$ at $d=5$. Note that the absolute
difference (the remainder term) changes only slightly with $\lambda$ in comparison to the percentage difference.  
\begin{table}[ht]
\begin{center}
\caption{Formula versus Numerical} 
\includegraphics[scale=0.65]{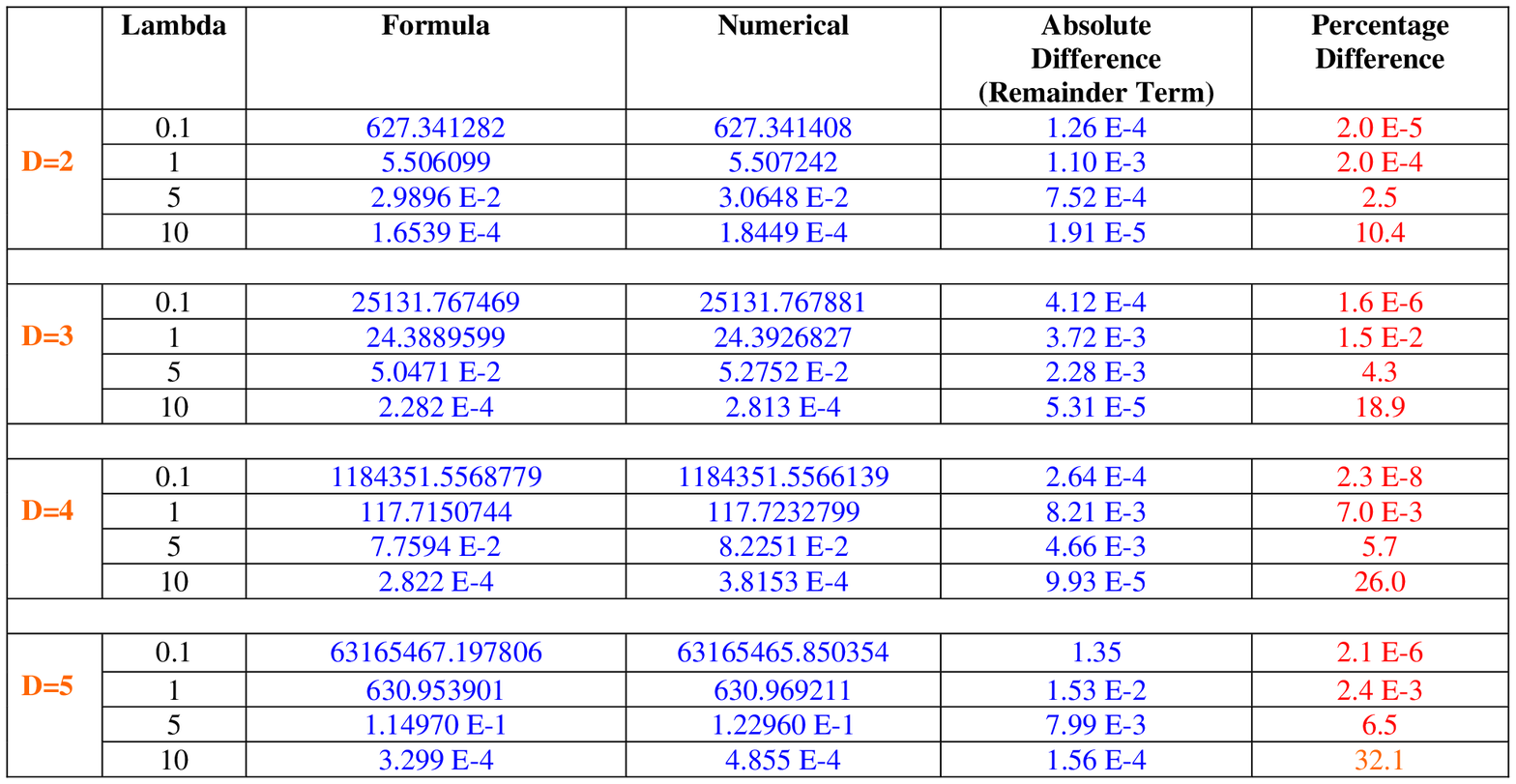}
\end{center}
\end{table}

\subsection*{\center{III. Casimir Energy and the Generating Function of $rd(n)$}}

Formula \reff{xid} for the generating function $\xi_d(\lambda)$ of $rd(n)$ has a direct relationship to the Casimir energy of a massless scalar field confined to a $d$-dimensional box of side $L$. In fact, the natural division of $\xi_d(\lambda)$ in two terms, $I_d(\lambda)$ and $C_d(\lambda)$, is `tailor made' for extracting the Casimir energy. Simply put, the expressions for the regularized vacuum energy with and without boundaries are expressed in terms of $\xi_d(\lambda)$ and $I_d(\lambda)$ repectively. The Casimir energy is their difference and is written in terms of $C_d(\lambda)$.

Consider a massless scalar field $\phi(x)$ confined to a box of side $L$ with Neumann boundary conditions where the derivative of the scalar field vanishes at the boundaries i.e. $\partial_i\phi(x) =0$ when $x_i =0$ or $L$. We now quickly go over the standard steps leading to to the expression for the regularized vacuum energy. The vacuum energy is the sum over all modes of $\tfrac{1}{2}\,\omega$ (we work in units where $\hbar=c=1$). For a massless scalar field confined to a $d$-dimensional box of side $L$ with Neumann boundary conditions, the frequency $\omega$ is given 
by 
\beq
\omega = \dfrac{\pi}{L} (n_1^2 + n_2^2 +\cdot + n_d^2)^{1/2}
\eeq{omega}
where $n_i \ge 0$. The vacuum energy is 
\beq
E_V = \dfrac{\pi}{2\,L} \sum_{n_d=0}^{\infty}\cdots\sum_{n_1=0}^{\infty}\,(n_1^2 + n_2^2 +\cdot + n_d^2)^{1/2} 
\eeq{vac}
which is formally infinite. We regularize it with an exponential term so that the regularized vacuum energy is
\beq
H_d(\lambda) = \dfrac{\pi}{2\,L} \sum_{n_d =0}^{\infty}\cdots\sum_{n_1=0}^{\infty}\,(n_1^2 + n_2^2 +\cdot + n_d^2)^{1/2} \,\expo{-\lambda\,(n_1^2 +n_2^2 + \cdots + n_d^2)^{1/2}}\,.
\eeq{reg}
We will later take the limit as $\lambda \to 0$. For now, notice the resemblance between $H_d(\lambda)$ and $-\partial_{\lambda} \,\xi_d(\lambda)$. Of course, the two are not equal since the sum in $\xi_d(\lambda)$ goes from $-\infty$ to $\infty$. However, we can express the $d$ sums from $0$ to $\infty$ found in $G_d(\lambda)$ in terms of sums from $-\infty$ to $\infty$. The function we are summing in \reff{reg} has the property $f(n) = f(-n)$ where $n$ can be $n_1,n_2$, etc. We therefore have the relation $\sum_{0}^{\infty} f(n) = \tfrac{1}{2}\,\sum_{-\infty}^{\infty} f(n) + \tfrac{1}{2} \,f(0)$
which can be expressed as an operator $\sum_{0}^{\infty} \to \tfrac{1}{2} \bigl( \sum_{-\infty}^{\infty} + 1\bigr)$. Applying the operator $d$ times yields 
\beq
\Bigl(\sum_{0}^{\infty}\Bigr)^d \to \dfrac{1}{2^d} \Bigl( 1+ \sum_{-\infty}^{\infty} \Bigr)^d =
\dfrac{1}{2^d} \sum_{p=1}^d \bino{d}{p} \Bigl(\sum_{-\infty}^{\infty}\Bigr)^p 
\eeq{operator}
where the case $p=0$ yields zero i.e. $f(n_1=0,n_2=0,...,n_d=0)=0$. We can therefore express \reff{reg} as 
\beq
H_d(\lambda) = -\dfrac{\pi}{L}\dfrac{1}{2^{d+1}}  \sum_{p=1}^d \bino{d}{p} \,\partial_{\lambda} \,\xi_p(\lambda) \,.
\eeq{yupie}
Equation \reff{yupie} shows already a connection between the generating function of $rd(n)$ and the regularized vacuum energy. We now want to calculate the Casimir energy. The Casimir energy is the difference between the vacuum energy with boundaries and the vacuum energy without boundaries. $H_d(\lambda)$ represents the energy with boundaries. Without boundaries the fields are continuous: there are no discrete modes.  This is equivalent to replacing sums by integrals. We have already seen that $\xi_p(\lambda) \to I_p(\lambda) -1$ when sums are replaced by integrals and hence, $\partial_{\lambda} \,\xi_p(\lambda) \to \partial_{\lambda} \,I_p(\lambda)$. So the regularized energy without boundaries, $G_d(\lambda)$, is
\beq
G_d(\lambda)= -\dfrac{\pi}{L}\dfrac{1}{2^{d+1}} \sum_{p=1}^d \bino{d}{p} \,\partial_{\lambda} \,I_p(\lambda) \,.
\eeq{e2d}
To return to the original vacuum energy we need to take the limit as $\lambda \to 0$. In this limit, both $G_d(\lambda)$ and $H_d(\lambda)$ are divergent. However, their difference in this limit is finite and equal to the Casimir energy $E_d$,
\beq
\begin{split}
E_d &\equiv \lim_{\lambda \to 0}  \bigl(H_d(\lambda) - G_d(\lambda)\bigr) =  -\dfrac{\pi}{L}\dfrac{1}{2^{d+1}}  \sum_{p=1}^d \bino{d}{p} \,\lim_{\lambda \to 0}\,\partial_{\lambda} \bigl(\xi_p(\lambda) -I_p(\lambda)\bigr)\\
&= -\dfrac{\pi}{L}\dfrac{1}{2^{d+1}} \sum_{p=1}^d \bino{d}{p} \,\lim_{\lambda \to 0} \,\partial_{\lambda}\, C_p(\lambda)\,.
\end{split}
\eeq{cdsf}
where \reff{xinew} was used. As we mentioned previously, $C_p(\lambda)$ is the only $\lambda-$-dependent term that finally enters into the calculation of the Casimir energy. From the definition \reff{IdCd} of  $C_d(\lambda)$ we obtain
\beq
\begin{split}
\lim_{\lambda \to 0} \, \partial_{\lambda}\,C_p(\lambda)&= \sum_{j=0}^{p-1} 2^{j+2}\,\,\Gamma(\tfrac{j+2}{2})\,\pi^{\frac{j}{2}}\,\lim_{\lambda \to 0} \chi_j(\lambda) \\
&= \sum_{j=0}^{p-1} \Gamma(\tfrac{j+2}{2})\,\pi^{\frac{-j-4}{2}}\, \zeta(j+2)
\end{split}
\eeq{dcd}
where we used the power series \reff{finalchi} for $\chi_j(\lambda)$ to obtain 
\beq
\lim_{\lambda \to 0} \chi_j(\lambda) = \dfrac{\zeta(j+2)}{(2\pi)^{j+2}}\,. 
\eeq{chilam}
Substituting \reff{dcd} into \reff{cdsf} yields the Neumann Casimir energy for massless scalar fields in a box of $d$ dimensions:
\beq
E_d = -\dfrac{1}{L}\dfrac{1}{\pi\,2^{d+1}} \sum_{p=1}^d \,\sum_{j=0}^{p-1} \bino{d}{p}\, \Gamma(\tfrac{j+2}{2})\,\pi^{\frac{-j}{2}}\, \zeta(j+2)\,.
\eeq{hgfrt}
Note that equation \reff{hgfrt} shows clearly that the Neumann Casimir energy $E_d$ is negative since the terms inside the sum are all positive. Values calculated using \reff{hgfrt} are in agreement with previous values calculated for the Neumann energy using different techniques \cite{Wolfram,Caruso,Li}. Similar expressions to \reff{hgfrt} can be obtained for other boundary conditions such as Dirichlet boundary conditions. 

\subsection*{\center{Acknowledgments}}

I wish to thank the Theoretical Physics Division at the University of Montreal for their hospitality while a visiting researcher during the summer of 2004 when part of this work was completed. I also thank Garnik Alexanian and Manu Paranjape for fruitful and lively discussions. This work was partly completed under Bishop's Senate Research Grant.

\begin{appendix}
\section{Appendix}
In this appendix we show that the remainder term $R_q$ is independent of the positive integers $q$ for the exponential function 
\beq
f(x) =\expo{-\lambda\,\sqrt{x^2 +C}}
\eeq{func}
where $C$ and $\lambda$ are positive real numbers. We want to show that $R_q=R_{q+1}$ where 
\beq
R_q = -\dfrac{1}{(2q)!} \int_0^1\, B_{2q}(x) \,\sum_{\nu=0}^{\infty}\, f^{2q}\,(x+\nu)\,dx\,. 
\eeq{Rq2}
The function $f(x)$ has the following properties:
\begin{itemize} 
\item odd derivatives evaluated at zero are zero i.e. $f^{2p-1}(0) =0$. This is proven in \cite{Ariel}.
\item the function and its derivatives are zero asymptotically \\i.e. $\lim\displaylimits_{x \to \infty} f(x) = 0$ and 
$\lim\displaylimits_{x \to \infty} f^p(x) = 0$. 
\end{itemize}
The Bernoulli functions have the following properties:
\begin{itemize} 
\item $B^{'}_n(x) = \dfrac{d}{dx} B_n(x) = n\,B_{n-1}(x)$.
\item $B_n(1) = (-1)^n\,B_n(0)$.
\item $B_{2n+1}(0) =0$ except for $B_1(0)=-1/2$. 
\end{itemize}
We use the above properties to obtain the following equality:
\beq
\begin{split}
B_{2q}(x)\, f^{2q}\,(x+\nu) &= \dfrac{B^{'}_{2q+1}(x)}{2q+1} \,f^{2q}\,(x+\nu)\\
&=\dfrac{d}{dx}\lrack \dfrac{B_{2q+1}(x)}{2q+1}\,f^{2q}\,(x+\nu)\rrack - \dfrac{B_{2q+1}(x)}{2q+1}\,f^{2q+1}\,(x+\nu)\,.
\end{split}
\eeq{yes} 
The integral in \reff{Rq2} is then given by
\beq
\begin{split}
\int_0^1\, B_{2q}(x) f^{2q}\,(x+\nu)\,dx &= \dfrac{B_{2q+1}(x)}{2q+1}\,f^{2q}\,(x+\nu)\Biggr\rvert_0^1\\
&-  \dfrac{1}{2q+1}\int_0^1\,B_{2q+1}(x) \, f^{2q+1}\,(x+\nu)\,dx \\
&=-  \dfrac{1}{2q+1}\int_0^1\,B_{2q+1}(x) \, f^{2q+1}\,(x+\nu)\,dx 
\end{split}
\eeq{check}
where we used the property $B_{2q+1}(1) = B_{2q+1}(0) =0$. Substituting \reff{check} into \reff{Rq2} yields  
\beq
R_q= + \dfrac{1}{(2q+1)!} \int_0^1\, B_{2q+1}(x) \,\sum_{\nu=0}^{\infty}\, f^{2q+1}\,(x+\nu)\,dx \,.
\eeq{Rq3}
We now repeat the process one more time. We obtain the equality  
\beq
\begin{split}
B_{2q+1}(x)\, f^{2q+1}\,(x+\nu) &= \dfrac{B^{'}_{2q+2}(x)}{2q+2} \,f^{2q+1}\,(x+\nu)\\
&=\dfrac{d}{dx}\lrack \dfrac{B_{2q+2}(x)}{2q+2}\,f^{2q+1}\,(x+\nu)\rrack \\&- \dfrac{B_{2q+2}(x)}{2q+2}\,f^{2q+2}\,(x+\nu)\,.
\end{split}
\eeq{yeh}  
The integral in \reff{Rq3} is then given by
\beq
\begin{split}
\int_0^1\, B_{2q+1}(x) f^{2q+1}\,(x+\nu)\,dx &= \dfrac{B_{2q+2}(x)}{2q+2}\,f^{2q+1}\,(x+\nu)\Biggr\rvert_0^1\\
&-  \dfrac{1}{2q+2}\int_0^1\,B_{2q+2}(x) \, f^{2q+2}\,(x+\nu)\,dx \\
\end{split}
\eeq{check2}
The first term yields, 
\beq
\begin{split}
\dfrac{B_{2q+2}(x)}{2q+2}\,f^{2q+1}\,(x+\nu)\Biggr\rvert_0^1 &= B_{2q+2}(1)f^{2q+1}\,(1+\nu) - B_{2q+2}(0)f^{2q+1}\,(\nu)\\
&=B_{2q+2}(0)\left(f^{2q+1}\,(1+\nu) - f^{2q+1}\,(\nu)\right)\,
\end{split}
\eeq{check3}
where we used the property $B_{2q+2}(1)=B_{2q+2}(0)$. Summing over $\nu$ yields,
\beq 
\begin{split}
\sum_{\nu=0}^{\infty}B_{2q+2}(0)&\left(f^{2q+1}\,(1+\nu) - f^{2q+1}\,(\nu)\right)\\
 &= B_{2q+2}(0)\left(\lim_{\nu\to\infty} f^{2q+1}\,(1+\nu) - f^{2q+1}\,(0)\right) = 0\,.
\end{split}
\eeq{zeros}
where we used the properties of $f(x)$ previously listed. Therefore, summing the first term over $\nu$ in \reff{check2} yields zero. Substituting \reff{check2} into \reff{Rq3} yields 
\beq
R_q= -\dfrac{1}{(2q+2)!} \int_0^1\, B_{2q+2}(x) \,\sum_{\nu=0}^{\infty}\, f^{2q+2}\,(x+\nu)\,dx
\eeq{Rq4}
which is the desired result. Since $R_q = R_{q+1}$, the remainder term is independent of $q$. 
\section{Appendix}
In this appendix we show that the term, $2^d\,\pi^{\frac{d-1}{2}}\, \Gamma(\tfrac{d+1}{2})/\lambda^d$, which appears in the formula \reff{xid}, is a $d$-dimensional volume integral obtained by replacing the sums by integrals in \reff{start}. Our goal is to show that
\beq 
I_d(\lambda) \equiv \int_{-\infty}^{\infty}\expo{-\lambda\,(x_1^2 +x_2^2 +\cdots +x_d^2)^{1/2}}\,dx_1\,dx_2\,\ldots dx_d = \dfrac{2^d}{\lambda^d}\,\,\pi^{\frac{d-1}{2}}\,\, \Gamma(\tfrac{d+1}{2}).
\eeq{app}
We switch from cartesian to spherical coordinates $(r,\phi,\theta_1,\theta_2,\ldots,\theta_{d-2})$
where $r =(x_1^2 + x_2^2 +\cdots + x_d^2)^{1/2}$. The $d$-dimensional volume element $dV$ in spherical coordinates is
\beq
\begin{split}
dV &= r^{d-1} \,dr \,d\phi\,\sin\,\theta_1\,d\theta_1\,\sin^2\,\theta_2\,d\theta_2\ldots
\sin^{d-2}\,\theta_{d-2}\,d\theta_{d-2} \\
&= r^{d-1} \,dr \,d\phi\prod_{k=1}^{d-2} \sin^k\,\theta_k\,d\theta_k 
\end{split}
\eeq{volume}
where $2\,\pi\!>\!\phi\!>\!0$, $\pi\!>\!\theta_i \!>\!0$ and $r$ runs from $0$ to $\infty$. Substituting \reff{volume} into \reff{app} yields
\beq
I_d(\lambda) =  2\,\pi \int_{0}^{\infty} \expo{-\lambda\,r}\, r^{d-1} \,dr \,\int_{0}^{\pi}\,\prod_{k=1}^{d-2} \sin^k\,\theta_k\,d\theta_k 
\eeq{integral}
where the $2\,\pi$ from the integral over $\phi$ was factored out. The integral over any $\theta$ yields
\beq
\int_{0}^{\pi} \sin^k\,\theta_k\,d\theta_k = \sqrt{\pi} \,\dfrac{\Gamma(\tfrac{k+1}{2})}{\Gamma(\tfrac{k+2}{2})}  
\eeq{thetaintegral}
and therefore
\beq
\int_{0}^{\pi}\,\prod_{k=1}^{d-2} \sin^k\,\theta_k\,d\theta_k = \dfrac{\pi^{\frac{d-2}{2}}}{\Gamma(d/2)}\,.
\eeq{gammaintegral} 
The integral over $r$ yields
\beq
\int_{0}^{\infty} \expo{-\lambda\,r}\, r^{d-1} \,dr =  \dfrac{\Gamma(d)}{\lambda^d}\,.
\eeq{rintegral}
Substituting \reff{rintegral} and \reff{gammaintegral} into \reff{integral} yields
\beq
I_d(\lambda) = 2 \,\dfrac{\pi^{\frac{d}{2}}}{\Gamma(d/2)}\dfrac{\Gamma(d)}{\lambda^d}\,.
\eeq{closeit}
The above expression can be simplified if we use the identity
\beq
\dfrac{\Gamma(d)}{\Gamma(d/2)} = \dfrac{1}{2}\,\dfrac{2^{\,d}\,\Gamma(\tfrac{d+1}{2})}{\sqrt{\pi}}\,.
\eeq{GammaIdentity}
Substituting \reff{GammaIdentity} into \reff{closeit} yields our desired result
\beq
I_d(\lambda) =\dfrac{2^d}{\lambda^d}\,\,\pi^{\frac{d-1}{2}}\,\, \Gamma(\tfrac{d+1}{2}) \,.
\eeq{realfinalo}
\end{appendix}

\end{document}